\documentclass[12pt,cite,epsf,epsfig]{article}
\usepackage{epsfig}

\begin{document}
\author{S. Dev\thanks{dev5703@yahoo.com} $^{,1}$,
Sanjeev Kumar\thanks{sanjeev3kumar@gmail.com} $^{,2}$, Surender
Verma\thanks{ s\_7verma@yahoo.co.in} $^{,1}$, Shivani
Gupta\thanks{shivani@prl.res.in} $^{,3}$\\ and R. R. Gautam\thanks{gautamrrg@gmail.com} $^{,1}$}
\title{Four Zero Texture Fermion Mass Matrices in SO(10) GUT}
\date{$^1$\textit{Department of Physics, Himachal Pradesh University, Shimla- 171005, INDIA.}\\
\smallskip
$^2$\textit{Department of Physics and Astrophysics, University of Delhi, Delhi- 110007, INDIA.}\\
\smallskip
$^3$\textit{Physical Research Laboratory, Navarangpura, Ahmedabad - 380009, INDIA}}

\maketitle
\begin{abstract}
We attempt the integration of the phenomenologically successful four zero texture of fermion mass matrices with the renormalizable SO(10) GUT. The resulting scenario is found to be highly predictive. Firstly, we examine the phenomenological implications of a class of the lepton mass matrices with parallel texture structures and obtain interesting constraints on the parameters of the charged lepton and the neutrino mass matrices. We combine these phenomenological constraints with the constraints obtained from SO(10) GUT to reduce the number of the free parameters and to further constrain the allowed ranges of the free parameters. The solar/atmospheric mixing angles obtained in this analysis are in fairly good agreement with the data. 
\end{abstract}
\section{Introduction}
The origin of the fermion masses and mixings along with the related problem of $CP$ violation constitute a formidable challenge for elementary particle physics. Leaving apart extremely small neutrino masses, even the charged fermion mass hierarchy ranges over at least five orders of magnitude. Since the fermion masses are derived from the Yukawa couplings, which are free parameters within the Standard Model (SM), these Yukawa couplings must span several orders of magnitude to accommodate the strongly hierarchical pattern of the fermion masses and mixings. However, the currently available data on the fermion masses and mixings is insufficient for an unambiguous reconstruction of the fermion mass matrices. To make matters worse, radiative corrections can obscure the underlying structures. Thus, the existing data cannot, without some additional assumptions, determine all the elements of the Yukawa coupling matrices for the quarks and the leptons. Some of these assumptions, invoked to restrict the form of the fermion mass matrices include the presence of texture zeros \cite{1}, requirement of zero determinant \cite{2} and vanishing minors \cite{3} to name just a few. The main motivation for invoking different mass matrix Ans\"{a}tze is to relate the fermion masses and mixing angles in a testable manner which reduces the number of free parameters in the Yukawa sector. The recent evidence for non-zero neutrino masses and mixings leads to a further proliferation of free parameters in the Yukawa sector. In the absence of a significant breakthrough in the theoretical understanding of the fermion flavors, the phenomenological approaches are bound to play a crucial role in interpreting new experimental data on the quark and the lepton mixing. These approaches are expected to provide useful hints towards unravelling the dynamics of the fermion mass generation, $CP$ violation and identification of possible underlying symmetries of the fermion flavors from which viable models of the fermion mass generation and flavor mixing could, hopefully, be constructed.\\
The strong fermion mass hierarchy should be apparent in the fermion mass matrices themselves with the contribution of smaller elements to physical masses and mixing angles expected to be negligibly small. Thus, these elements can, effectively, be neglected and replaced by zeros: the so-called texture zeros. However, the current neutrino oscillation data is consistent only with a limited number of texture schemes \cite{1}. Specifically, the available neutrino oscillation data disallow all the neutrino mass matrices with three or more texture zeros \cite{1} in the flavor basis. The texture zeros at different positions in the neutrino mass matrix, in particular, and the fermion mass matrices, in general, could be the consequence of some underlying symmetry. Such universal textures of the fermion mass matrices can be realized within the framework of Grand Unified Theories (GUTs). Though Grand Unification on its own does not shed any light on the flavor problem, the GUTs provide the optimal framework in which possible solutions to the flavor problem could be embedded. Grand unified models attempt to explain the masses and mixings in both the quark and the lepton sectors simultaneously. The textures for the mass matrices obtained in these models can either be assumed at the very outset or can be derived from the observed mixing matrix in the flavor basis. Alternatively, the textures of the mass matrix can be obtained by embedding some family symmetry within the chosen Grand Unification group. One particularly interesting class of models is that based upon the SO(10) Grand Unification group. There are two kinds of minimal models in this class: those based upon Higgs dimension \textbf{10, 126}, $\overline{\textbf{126}}$ and possibly also \textbf{120} and/or \textbf{210} and those based upon \textbf{10, 16,} $\overline{\textbf{16}}$ and \textbf{45} representations. The former choice, generally, has symmetric and/or antisymmetric texture mass  matrices while the latter type generally imply lopsided mass matrices for the down type quarks and the charged leptons. In the present work, we consider the four zero texture [FTZ] Ans\"{a}tz for the fermion mass matrices within the SO(10) GUT framework. Within this framework, we not only have a relation between the down type quark mass matrices ($M_d$) and the charged lepton mass matrices ($M_l$) but also a relation between the up-type quark mass matrices ($M_u$) and the Dirac neutrino mass matrices ($M_D$).

Four zero texture Ans\"{a}tz is especially important since it can successfully describe not only the quark but also the lepton sector including the charged lepton and the neutrino masses. Moreover, these are compatible with specific GUT models \cite{4} and can be obtained from Abelian flavor symmetries \cite{5}. Furthermore, these mass matrices can accommodate the present value of Sin$2\beta$ \cite{6}. In view of the phenomenological success of the FZT Ans\"{a}tz, it would be interesting to examine it in the larger context of SO(10) GUT. The FZT Ans\"{a}tz has been studied earlier \cite{7} in the context of SO(10) GUT which, in general, leads to relations between mass matrices. On the other hand texture zeros imply relations between the elements of the mass matrices. It is, therefore, interesting to examine FZT Ans\"{a}tze within the framework of SO(10) GUT. In the earlier analysis \cite{7} somehow, the constraints implied by texture zeros of the neutrino mass matrix have not been incorporated into the analysis and only SO(10) constraints on mass matrices have been used. In the present work, we have obtained the FZT constraints on the lepton mass matrices in Sec. 2. These constraints have been incorporated into the analysis of SO(10) GUT with four zero texture structure in Sec. 3. We have also presented the Takagi diagonalization of the general complex symmetric neutrino mass matrix in Sec. 3 without making any simplifying assumption for the phases. Conclusions have been summarized in Sec. 4.
\section{Four zero texture lepton mass matrices}
All information about the lepton masses and mixings is encoded in a hermitian charged lepton mass matrix $M_e$ and the neutrino mass matrix $M_{\nu}$ which is complex symmetric. Both these matrices have parallel texture structure i.e. zeros at (1,1) and (1,3) places \cite{8}. In the present work, we consider a special case of FZT Ans\"{a}tze in which the lepton mass matrices have parallel texture structure with zero entries at (1, 1) and (1, 3) places. 
The charged fermion mass matrices are assumed to be of the form
\begin{equation}
M_f=\left(%
\begin{array}{ccc}
  0 & a_f & 0 \\
  a_f^* & b_f & c_f \\
  0 & c_f^* & d_f \\
\end{array}%
\right)
\end{equation}
where $f=e,u$ and $d$ for charged lepton, up-type quarks and down-type quarks, respectively. The elements $a_f$ and $c_f$ are complex with $a_f=|a_f|e^{i\phi_f}$, $c_f=|c_f|e^{i\psi_f}$. The elements $b_f$ and $d_f$ are real because of hermiticity of $M_f$.
The unitary matrices $V_f$ diagonalize the mass matrices $M_f$:
\begin{equation}
M_f=V_fM_f^dV_f^\dagger.
\end{equation}
Here $M_{f}^d=(-m_{f1},m_{f2},m_{f3})$ and $V_f=P_fO_f$ where $P_f=diag(e^{i\phi_f},1,e^{i\psi_f})$. $O_f$ is the diagonalization matrix for the real symmetric mass matrix $M_f^{(r)}=P_f^{\dagger}M_fP_f$. Using the invariants, Tr$M_f^{r}$, Tr$M_f^{r^ 2}$ and Det$M_f^{r}$ three elements $(a_f, b_f, c_f)$ of the total four elements can be written in terms of remaining one element $d_f$ and the mass eigenvalues $-m_{f1}$, $m_{f2}$ and $m_{f3}$ of the fermion mass matrix $M_f$ as
\begin{eqnarray}
b_f=-m_{f1}+m_{f2}+m_{f3}-d_f,\nonumber\\
|a_f|=\left(\frac{m_{f1}m_{f2}m_{f3}}{d_f}\right)^{\frac{1}{2}},\\
|c_f|=\left[-\frac{(d_f+m_{f1})(d_f-m_{f2})(d_f-m_{f3})}{d_f}\right]^{\frac{1}{2}}.\nonumber
\end{eqnarray}  
The free parameter $d_f$ should be in the range
$m_{f2}<d_f<m_{f3}$ for the elements $|a_f|$ and
$|c_f|$ to be real. Using Eqn.(3) the elements of the diagonalization matrix, $O_f$ can be written in terms of the charged fermion masses and the charged fermion mass matrix element $d_f$ as
\begin{equation}
 \left.\begin{array}{l}

O_{f11}=\sqrt{\frac{m_{f2}m_{f3}(d_f+m_{f1})}{d_f
(-m_{f1}-m_{f2})(-m_{f1}-m_{f3})}}
\\

O_{f12}=\sqrt{\frac{-m_{f1}m_{f3}(d_f-m_{f2})}{d_f
(m_{f2}-m_{f3})(m_{f2}+m_{f1})}}   \\

O_{f13}=\sqrt{\frac{-m_{f1}m_{f2}(d_f-m_{f3})}{d_f
(m_{f3}-m_{f2})(m_{f3}+m_{f1})}}   \\

O_{f21}=-\sqrt{\frac{m_{f1}(d_f+m_{f1})}{(-m_{f1}-m_{f2})(-m_{f1}-m_{f3})}}   \\

O_{f22}=\sqrt{-\frac{m_{f2}(d_f-m_{f2})}{
(m_{f2}-m_{f3})(m_{f2}+m_{f1})}} \\

O_{f23}=\sqrt{-\frac{m_{f3}(d_f-m_{f3})}{
(m_{f3}-m_{f2})(m_{f3}+m_{f1})}} \\

O_{f31}=\sqrt{\frac{-m_{f1}(d_f-m_{f2})(d_f-m_{f3})}{d_f
(-m_{f1}-m_{f2})(-m_{f1}-m_{f3})}}  \\

O_{f32}=-\sqrt{\frac{m_{f2}(d_f-m_{f3})(d_f+m_{f1})}{d_f
(m_{f2}-m_{f3})(m_{f2}+m_{f1})}}  \\

O_{f33}=\sqrt{\frac{m_{f3}(d_f+m_{f1})(d_f-m_{f2})}{d_f
(m_{f3}-m_{f2})(m_{f3}+m_{f1})}} 
\end{array}  \right\}.
\end{equation}
The neutrino mass matrix
\begin{equation}
M_\nu=\left(%
\begin{array}{ccc}
  0 & a_\nu & 0 \\
  a_\nu& b_\nu & c_\nu \\
  0 &c_\nu  & d_\nu \\
\end{array}
\right)
\end{equation}
is complex symmetric for which all the four elements are in general complex and is diagonalized by a complex unitary matrix $V_{\nu}$:
\begin{equation}
M_\nu=V_\nu M_ \nu^{diag}V_\nu^T.
\end{equation}
where $M_\nu^{diag}=diag(-m_1,m_2,m_3)$
The diagonalization matrices $V_e$, $V_\nu$, $V_u$ and $V_d$ are not physically observable but the products
$U_{CKM}=V_u^\dagger V_d$ \cite{9}, and 
$U_{PMNS}=V_e^\dagger V_\nu$ \cite{10}
are the physically observable quark and lepton mixing matrices, respectively. They can be parametrized in terms of the mixing angles and $CP$ violating phases. Considering neutrinos to be the Majorana particles, we parametrize $U_{PMNS}=U.P$ in terms of the three mixing angles ($\theta_{12}$, $\theta_{23}$ and $\theta_{13}$) and the three $CP$ violating phases ($\alpha$, $\beta$ and $\delta$) as
\begin{equation}
U_{PMNS}=\left(
\begin{array}{ccc}
c_{12}c_{13} & s_{12}c_{13} & s_{13}e^{-i\delta } \\
-s_{12}c_{23}-c_{12}s_{23}s_{13}e^{i\delta } &
c_{12}c_{23}-s_{12}s_{23}s_{13}e^{i\delta } & s_{23}c_{13} \\
s_{12}s_{23}-c_{12}c_{23}s_{13}e^{i\delta } &
-c_{12}s_{23}-s_{12}c_{23}s_{13}e^{i\delta } & c_{23}c_{13}
\end{array}
\right) \left(
\begin{array}{ccc}
1 & 0 & 0 \\
0 & e^{i\alpha } & 0 \\
0 & 0 & e^{i\left( \beta +\delta \right) }
\end{array}
\right)
\end{equation}
where $s_{ij}=\sin\theta_{ij}$ and $c_{ij}=\cos\theta_{ij}$ for $i,j$=1, 2, 3. 

The (1,1) and (1,3) entries of the neutrino mass matrix $M_{\nu}$ are assumed to be zero. Thus, we obtain two complex homogeneous linear equations for $m_1$, $\tilde{m_2}$ and $\tilde{m_3}$ given by
\begin{equation}
m_1a^2+\tilde {m_2}b^2+\tilde{m_3}c^2=0,
\end{equation}
\begin{equation}
m_1ad+\tilde {m_2}bg+\tilde{m_3}ch=0
\end{equation}
where the complex coefficients $a$, $b$, $c$, $d$, $g$ and $h$ are given by
\begin{equation}
\left.\begin{array} {c}
a=O_{e11}U_{e1}+O_{e12}U_{m1}+O_{e13}U_{t1},\nonumber\\
b=O_{e11}U_{e2}+O_{e12}U_{m2}+O_{e13}U_{t2},\nonumber\\
c=O_{e11}U_{e3}+O_{e12}U_{m3}+O_{e13}U_{t3},\nonumber\\
d=O_{e31}U_{e1}+O_{e32}U_{m1}+O_{e33}U_{t1},\nonumber\\
g=O_{e31}U_{e2}+O_{e32}U_{m2}+O_{e33}U_{t2},\nonumber\\
h=O_{e31}U_{e3}+O_{e32}U_{m3}+O_{e33}U_{t3},\nonumber\\
\end{array} \right\}
\end{equation}
and $\tilde{m_2}=m_2e^{2i\alpha}$, $\tilde{m_3}=m_3e^{2i(\beta+\delta)}$.
Solving Eqns.(8-9) for the two mass ratios
$\left(\frac{m_1}{m_2}\right)$ and $\left(\frac{m_1}{m_3}\right)$,
we obtain
\begin{equation}
\begin{large}
\begin{array}{c}
\frac{m_1}{m_2}e^{-2i\alpha}=\frac{b(cg-bh)}{a(ah-cd)},\\
\frac{m_1}{m_3}e^{-2i\beta}=\frac{c(bh-cg)}{a(ag-bd)}e^{2i\delta}.
\end{array}
\end{large}
\end{equation}
One can enumerate the number of parameters in Eqn.(11). The nine parameters [three neutrino mixing angles ($\theta_{12}$, $\theta_{23}$, $\theta_{13}$), three neutrino mass eigenvalues ($m_1$, $m_2$, $m_3$), two Majorana-type $CP$ violating phases ($\alpha$, $\beta$) and Dirac-type $CP$ violating phase, $\delta$] come from the neutrino sector and the four parameters [three charged lepton masses($m_e$, $m_{\mu}$, $m_{\tau}$) and free parameter $d_e$] come from the charged lepton sector, thus, totalling thirteen parameters. The three charged lepton masses are \cite{11}
\begin{equation}
m_e=0.510998910 MeV, \\
m_{\mu}=105.658367 MeV, \\
m_{\tau}=1776.84 MeV. \\
\end{equation}
The experimental constraints on the neutrino oscillation parameters with their best fit values and 3$\sigma$ deviations are \cite{12}
\begin{equation}
\begin{array}{c}
\Delta m_{12}^{2} =7.67_{-0.53}^{+0.52}\times 10^{-5}eV^{2},
 \\
\Delta m_{31}^{2} = 2.39_{-0.33}^{+0.42}\times 10^{-3}eV^{2},
 \\
\theta_{12}^o =33.95_{-3.1}^{+3.81},   \\
\theta_{23}^o =43.05_{-7.93}^{+10.32}.
\end{array}
\end{equation}

\begin{figure}
\begin{center}
{\epsfig{file=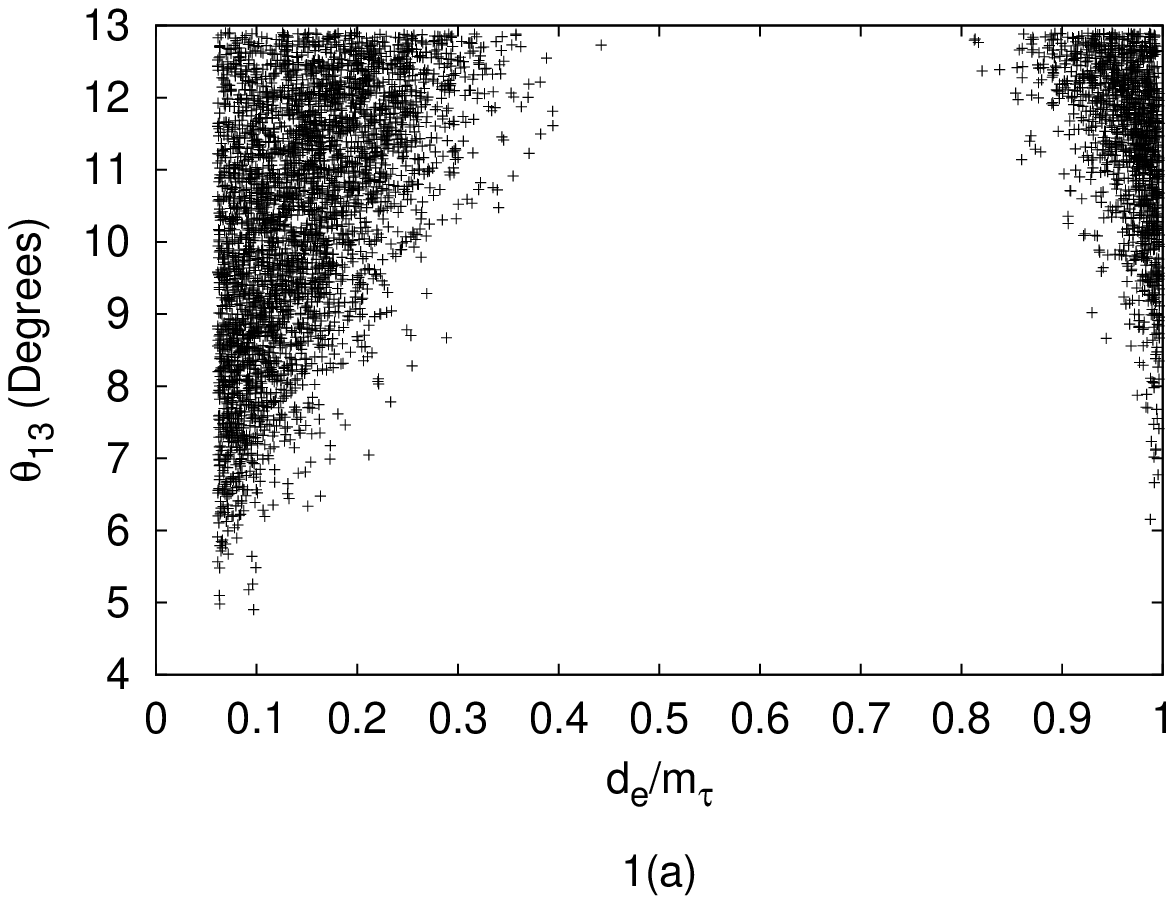, width=5.0cm, height=5.0cm}\epsfig{file=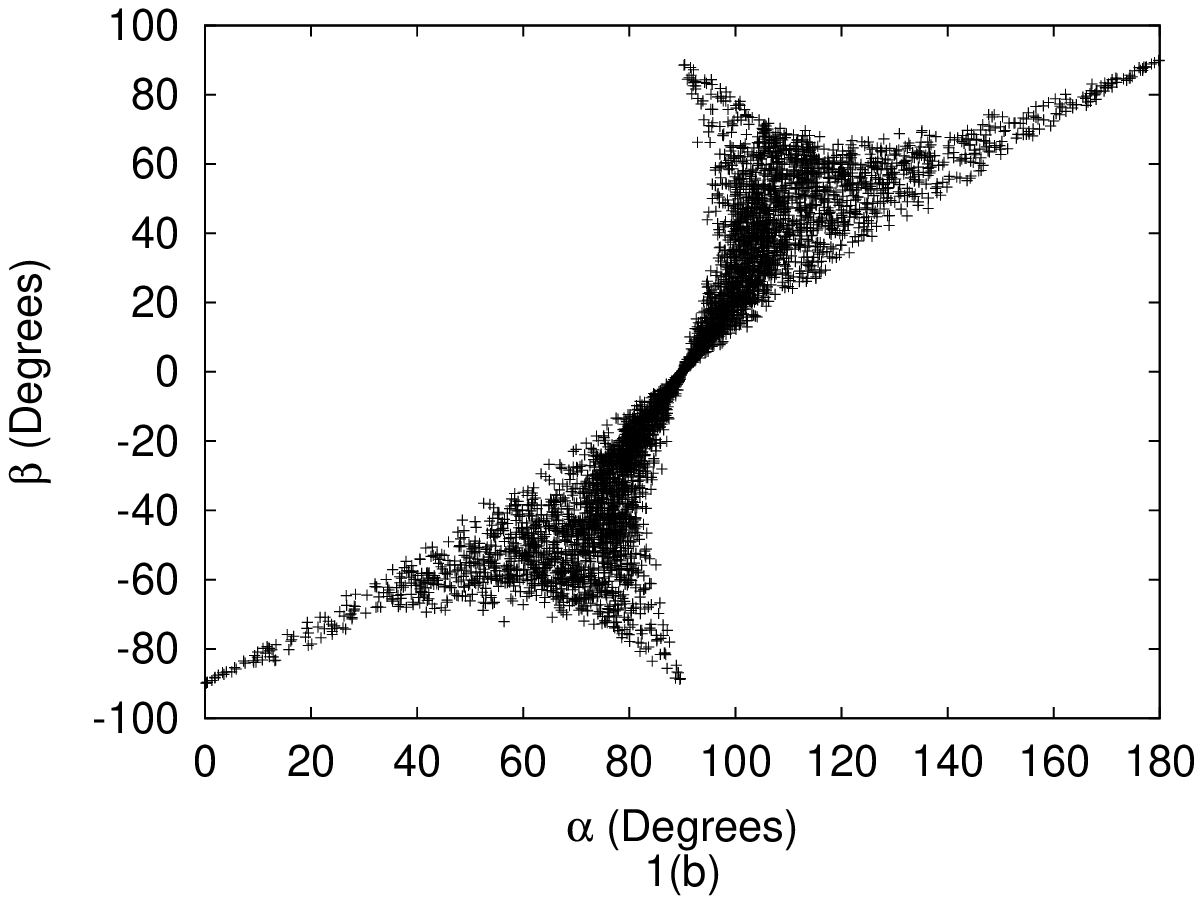, width=5.0cm, height=5.0cm}}\\
{\epsfig{file=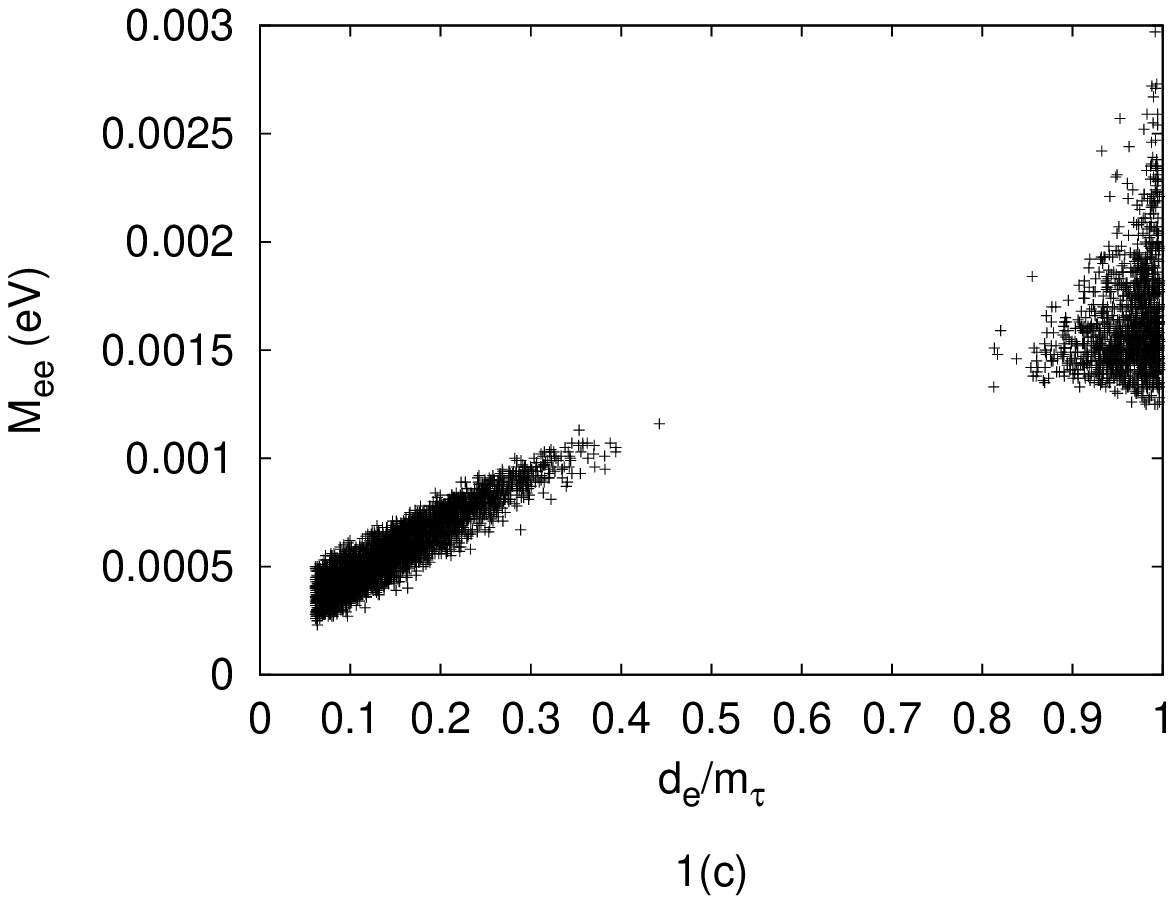, width=5.5cm, height=5.0cm}}
\end{center}
\caption{Correlation plots for $d_e/m_{\tau}$, $\theta_{13}$, $M_{ee}$ and the Majorana phases at $3\sigma$.}
\end{figure}
In addition, there is an upper bound on the mixing angle $\theta_{13}$
from the CHOOZ experiment($\theta_{12}^o<12.38$). Recently T2K\cite{13} and MINOS\cite{14} experiments have given hints of a relatively large $\theta_{13}$.\\
For the simultaneous existence of two texture zeros at $(1,1)$ and
$(1,3)$ positions in $M_{\nu}$, the two values of $m_1$ given by
\begin{equation}
m_{1}=\left|\frac{m_1}{m_2}\right| \sqrt{\frac{ \Delta
m_{12}^{2}}{1-\left|\frac{m_1}{m_2}\right| ^{2}}}
\end{equation}
and
\begin{equation}
m_{1}=\left|\frac{m_1}{m_3}\right| \sqrt{\frac{\Delta m_{12}^{2}+
\Delta m_{23}^{2}}{ 1-\left|\frac{m_1}{m_3}\right|^{2}}},
\end{equation}
calculated from the mass ratios
$\left(\frac{m_1}{m_2},\frac{m_1}{m_3}\right)$, respectively, must
be equal to within the errors of oscillation data. This constraint can be used to constrain the unknown parameters $\delta$, $d_e$ and the third mixing angle $\theta_{13}$ which has not been measured experimentally as yet. We, thus, obtain not only the correlations [Fig. 1(a), Fig. 1(b)] amongst the unknown parameters [viz. the neutrino mixing angles, the $CP$ violating phases and $d_e$] but also constrain the unknown parameter $d_e$. We obtain two allowed regions for the free parameter $d_e$. We call them `low $d_e$' and `high $d_e$' solutions. The other two free parameters ($\phi_e$ and $\psi_e$) are not constrained by the presence of texture zeros because they factor out from Eqns.(8) and (9). We have, also, calculated the effective Majorana mass $M_{ee}$ (Fig.1(c)) appearing in the
neutrinoless double beta decay for the allowed parameter space
\begin{equation}
M_{ee}=|m_1 U_{e1}^2+m_2 U_{e2}^2+m_3 U_{e3}^2|.
\end{equation}
These constraints on the hitherto unknown parameter $d_e$ obtained from the FZT Ans\"{a}tz will be used as an input in the next section where we attempt to constrain the SO(10) GUT. 
\section{Four Zero Texture and SO(10) GUT}
In the next step we combine the  FZT Ans\"{a}tz with the constraints obtained from the SO(10) GUT. Such an analysis will require renormalization group (RG) running from weak scale to the GUT scale. However, it is known that the effects of RG running are negligible for mass matrices with normal hierarchy. Having obtained the texture zero constraints on the free parameter $d_e$, we now further constrain it by imposing the following SO(10) GUT relations
\begin{equation}
\left.\begin{array} {l}
M_u=S+\delta^{''} A+ \epsilon S'\equiv  S_u+A_u, \\ \nonumber
M_d=\eta S+\delta{'} A+S' \equiv S_d+A_d, \\ \nonumber
M_D=S+\delta{'''} A-3\epsilon S'  \equiv S_D+A_D, \\ \nonumber
M_e=\eta S+A-3S'  \equiv S_e+A_e, \\ \nonumber
M_L=\rho S'  \equiv S_L, \\ \nonumber
M_R=\gamma S'\equiv S_R.
\end{array} \right\}
\end{equation}  
where $S$, $S'$ are the symmetric part coming from the \textbf{10} and \textbf{126} dimensional Higgs representations, respectively, and $A$ is the antisymmetric part coming from the \textbf{120} dimensional Higgs representation. Here, $\eta$, $\epsilon$, $\rho$, $\gamma$, $\delta{'}$, $\delta{''}$ and $\delta{'''}$ represent the relative coefficients of the vacuum expectation values (VEVs). From Eqn.(17), we find that eighteen parameters come from $S_f+A_f$, $f=u, d, e$ and four are the SO(10) coefficients $\delta{'}$, $\eta$, $\epsilon$ and $\delta{''}$, thus, totalling twenty two free parameters.
Eqn.(17) implies the following relations
\begin{eqnarray}
 4\eta S_u=(3+\eta\epsilon)S_d+(1-\eta\epsilon)S_e, \\ \nonumber
\delta{'} A_u=\delta{''} A_d=\delta{'}\delta{''} A_e.
\end{eqnarray}
In the component form, the above set of SO(10) constraints (Eqn.(17)) can be written as
\begin{equation}
\left.\begin{array} {l}
4 \eta a_u \cos{(\Delta \phi+\phi_d)}=(3+\eta\epsilon)a_d\cos{\phi_d}+(1-\eta\epsilon) a_e \cos{\phi_e}, \\ \nonumber
4 \eta c_u \cos{(\Delta \psi+\psi_d)}=(3+\eta\epsilon)c_d\cos{\psi_d}+(1-\eta\epsilon) c_e \cos{\psi_e}, \\ \nonumber
4 \eta b_u=(3+\eta \epsilon) b_d+(1-\eta\epsilon)b_e, \\ \nonumber
4 \eta d_u=(3+\eta\epsilon) d_d+(1-\eta\epsilon)d_e, \\ \nonumber
\delta{'} a_u\sin{(\Delta \phi+\phi_d)}=\delta{''} a_d \sin{\phi_d}=\delta{'}\delta{''} a_e \sin{\phi_e}, \\ \nonumber
\delta{'} c_u\sin{(\Delta \psi+\psi_d)}=\delta{''} c_d \sin{\psi_d}=\delta{'}\delta{''} c_e \sin{\psi_e}. \\ \nonumber
\end{array} \right\}
\end{equation}
In the above, we have largely followed the notations and conventions of Ref.[7] which combines the SO(10) constraints in the form of Eqn.(19).
We have 8 independent equations on 22 free parameters appearing in the quark and the charged lepton sectors. However, using the experimental results on the quark/charged lepton masses and mixings which includes six quark masses, three CKM mixing angles, one CKM $CP$ violating phase and three charged lepton masses, we are left with only one free parameter to be fixed. These eight independent equations can be used to relate $d_e$, $\phi_e$ and $\psi_e$ with the corresponding parameters in the quark sector i.e. $d_u$, $d_d$, $\Delta\phi=\phi_u-\phi_d$ and $\Delta\psi=\psi_u-\psi_d$ (we note that only these phase differences are determined from the experimental data). The phases $\phi_e$ and $\psi_e$ are given by the equations
\begin{equation}
\cos{\phi_e}\equiv\frac{4 \eta a_u \cos{(\Delta \phi+\phi_d)}-(3+\kappa)a_d\cos{\phi_d}}{(1-\kappa)a_e}, 
\end{equation}
\begin{equation}
\cos{\psi_e}\equiv\frac{4 \eta c_u \cos{(\Delta \psi+\psi_d)}-(3+\kappa)c_d\cos{\psi_d}}{(1-\kappa)c_e},
\end{equation}
where the down type quark phases are 
\begin{equation}
\tan{\phi_d}=\frac{a_u\sin{\Delta\phi}}{r a_d-a_u\cos{\Delta\phi}}, 
\end{equation}
\begin{equation}
\tan{\psi_d}=\frac{c_u\sin{\Delta\psi}}{r c_d-c_u\cos{\Delta\psi}},
\end{equation}
with $r=\frac{\delta{''}}{\delta{'}}$ and
\begin{equation}
\eta=\frac{(m_d+m_s+m_b)d_e-(m_e+m_{\mu}+m_\tau)d_d}{(m_d+m_s+m_b-m_e-m_{\mu}-m_{\tau})d_u-(m_u+m_c+m_t)(d_d-d_e)}, \\ \nonumber
\end{equation}
\begin{equation}
\kappa=\frac{(m_u+m_c+m_t)(3d_d+d_e)-(3(m_d+m_s+m_b)+(m_e+m_{\mu}+m_{\tau}))d_u}{(m_d+m_s+m_b-m_e-m_{\mu}-m_{\tau})d_u-(m_u+m_c+m_t)(d_d-d_e)}.
\end{equation}
The parameter $d_e$ is constrained by the requirement that $\cos\phi_e$ and $\cos\psi_e$ in Eqns.(20-21) are physical and 
\begin{equation}
\frac{c_d \sin\psi_d}{a_d \sin\phi_d}=\frac{c_e \sin\psi_e}{a_e \sin\phi_e}.
\end{equation}
\begin{figure}
\begin{center}
{\epsfig{file=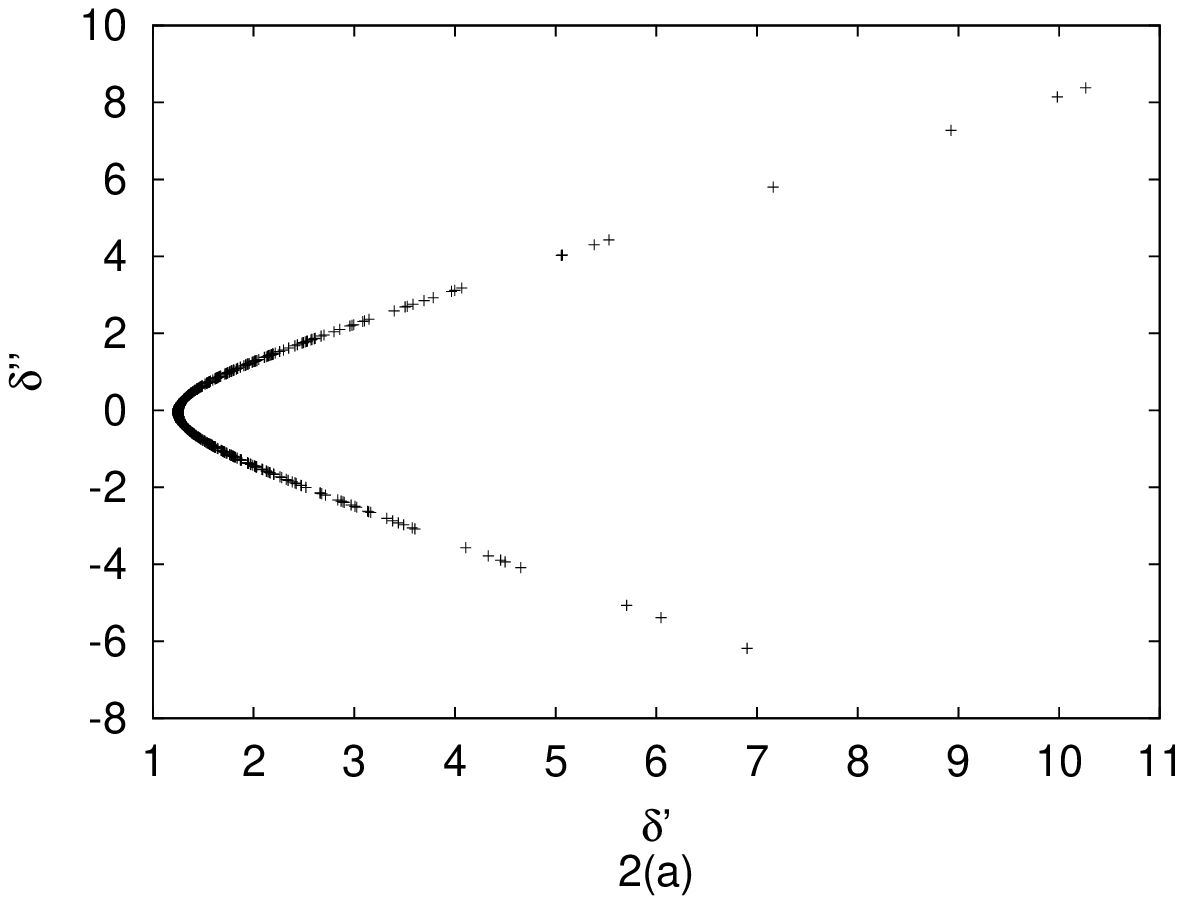, width=6.0cm, height=6.0cm}\epsfig{file=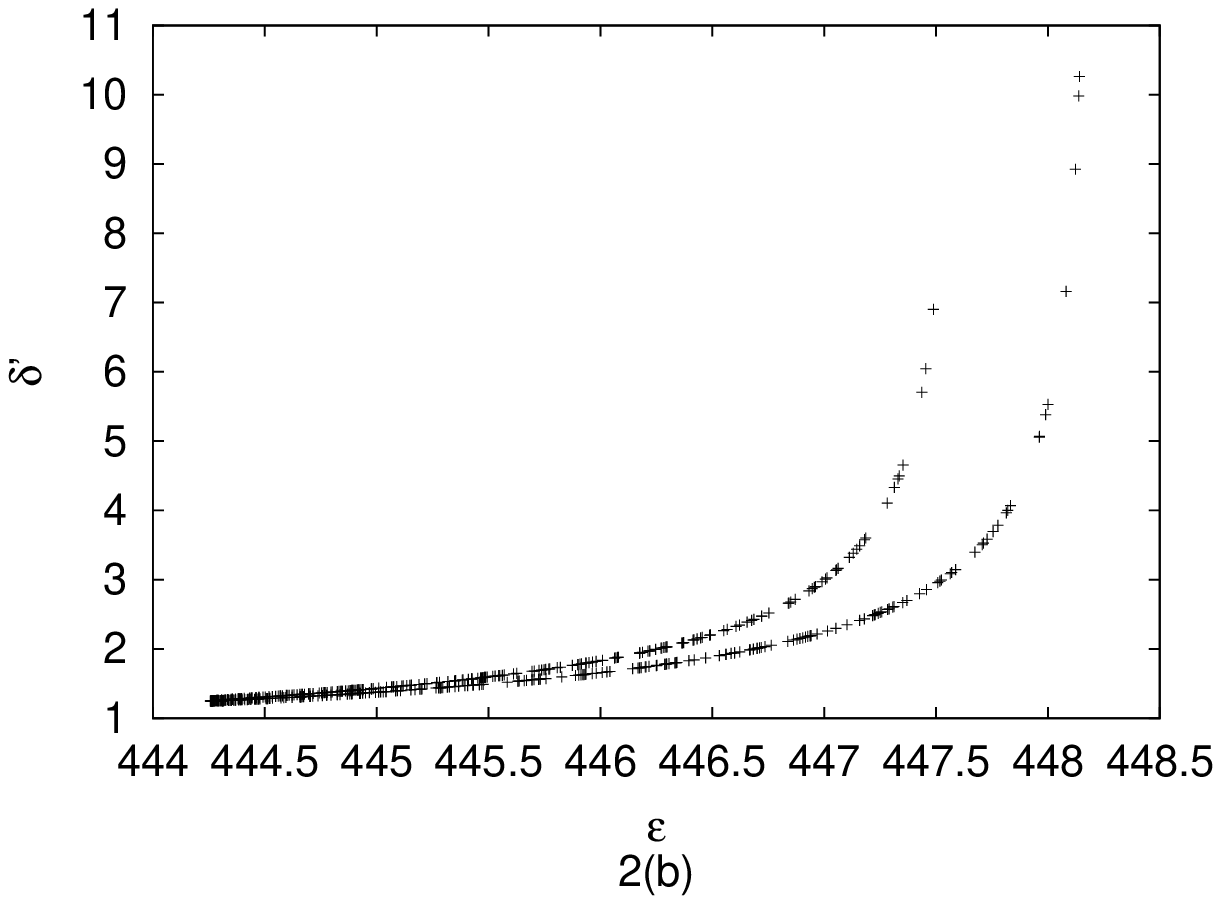, width=6.0cm, height=6.0cm}}\\
{\epsfig{file=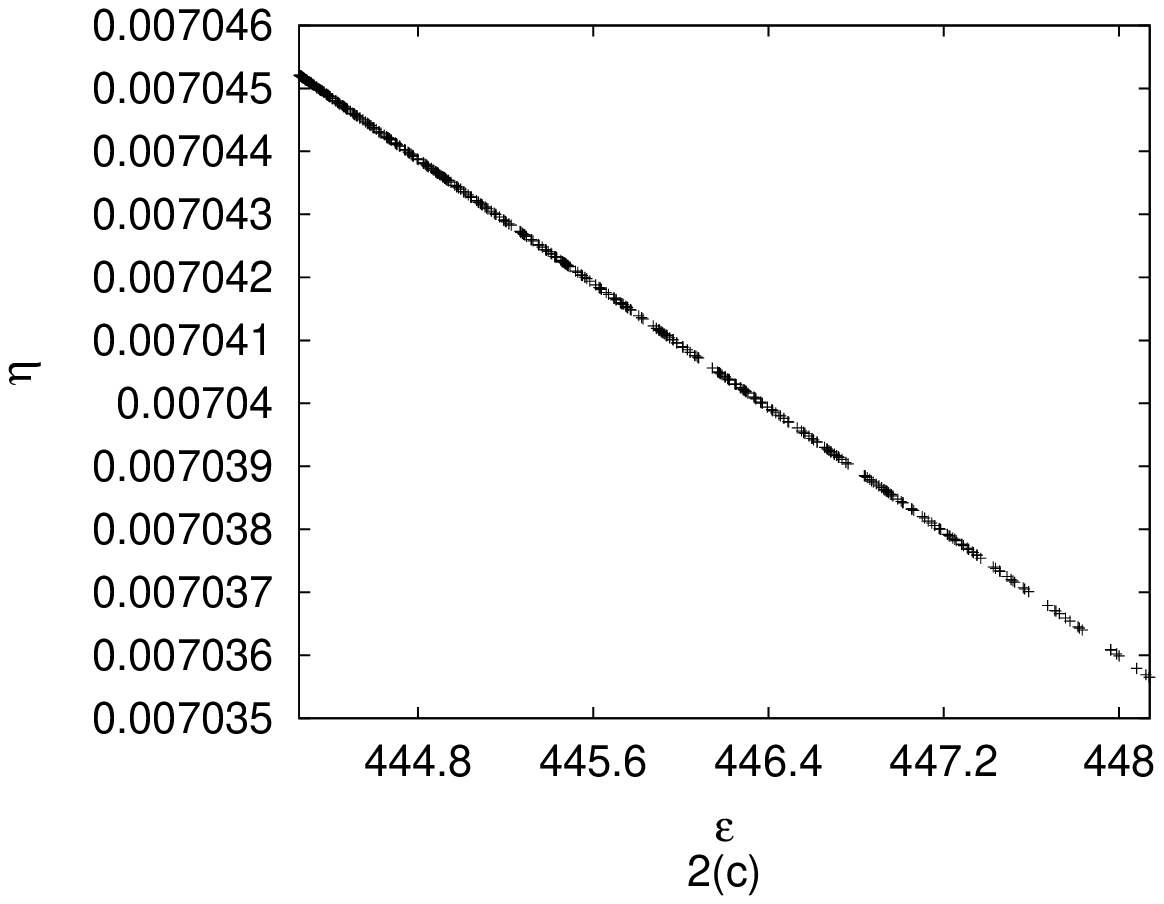, width=6.5cm, height=6.0cm}}
\end{center}
\caption{Correlation plots for some of the SO(10) parameters for high $d_e$.}
\end{figure}
The parameters $d_u$, $d_d$, $\Delta\phi$ and $\Delta\psi$ can be constrained by the observed CKM matrix \cite{6}. The best fit values realized for these parameters are given by
\begin{equation}
\Delta\phi=\frac{\pi}{2},
\end{equation}
\begin{equation}
\Delta\psi=-0.12,
\end{equation}
\begin{equation}
d_u=0.95 m_t, \\ \nonumber
\end{equation}
\begin{equation}
d_d=0.94m_b.
\end{equation} 
In our numerical analysis, we have used the following best fit values for the quark and the charged lepton masses \cite{7}:
\begin{equation}
\begin{array} {c}

m_u=1.04 MeV, m_c=302 MeV, m_t=129 GeV,\\ \nonumber
m_d=1.33 MeV, m_s=26.5 MeV, m_b=1.0 GeV,\\ \nonumber
m_e=0.325 MeV, m_{\mu}=68.55 MeV, m_{\tau}=1171.62 MeV. \\ \nonumber
\end{array}
\end{equation}
The correlation plots between $\phi_e$, $\psi_e$ and $d_e/m_\tau$ for the best fit values of $d_u$, $d_d$, $\Delta\phi$ and $\Delta\psi$ have been plotted in Fig.(3).
\begin{figure}
\begin{center}
{\epsfig{file=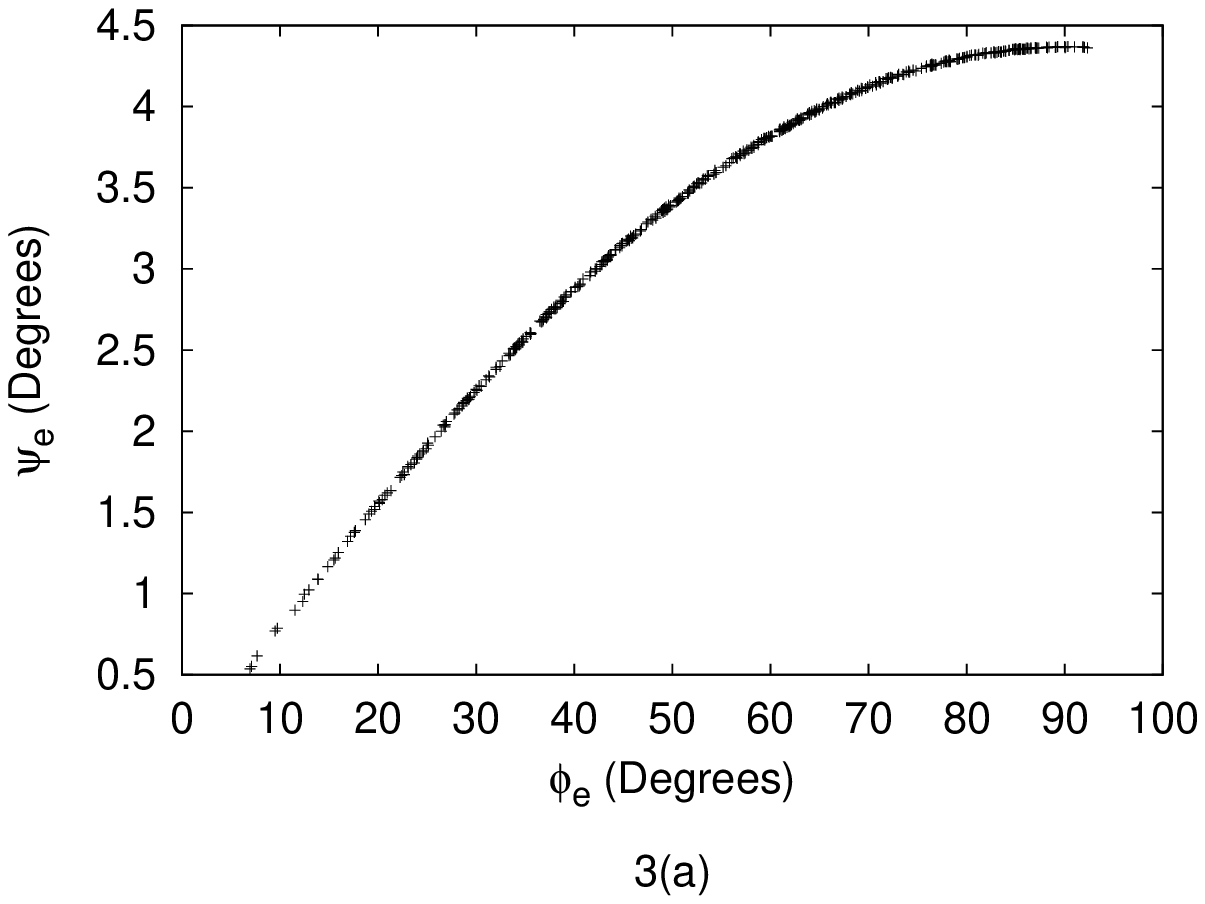, width=5.0cm, height=5.0cm}\epsfig{file=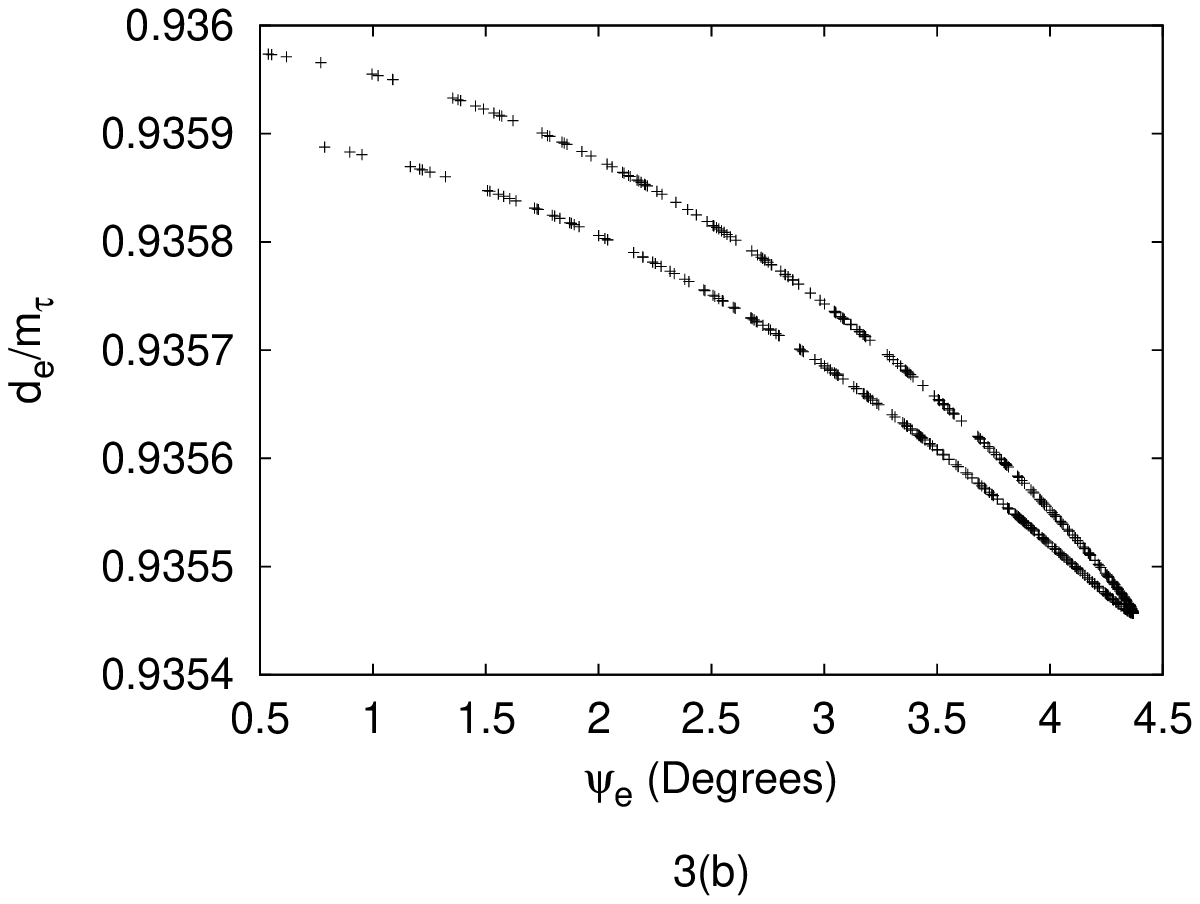, width=5.0cm, height=5.0cm}}\\
\end{center}
\caption{Correlation plots for phases from the charged lepton sector and $d_e/m_\tau$ (higher region).}
\end{figure}
In the phenomenological analysis of the FZT Ans\"{a}tz, presented in the previous section, we obtained two allowed regions for the free parameter $d_e$ compatible with the available data on lepton masses and mixings. Using these phenomenologically allowed ranges for $d_e$ along with the SO(10) GUT relations we constrain the $d_e-r$ plane. After the combined input of zero textures and SO(10) relations, the allowed ranges of $d_e$ are $(0.307346 m_\tau<d_e<0.309016 m_\tau)$ for the low $d_e$ and $(0.935427 m_\tau<d_e<0.935991 m_\tau)$ for high $d_e$. So far we have succeeded in fitting the unknown parameters of the charged lepton sector for the best fit values of $d_u$, $d_d$, $\Delta\phi$ and $\Delta\psi$. \\
The neutrino mass matrix from Type-I+II seesaw mechanism \cite{15} is given by
\begin{equation}
M_{\nu}=M_L-M_DM_R^{-1}M_D^T
\end{equation}
which introduces three more unknown parameters $\rho$, $\gamma$ and $\delta{'''}$ (Eqn.(17)). The lepton mixing matrix is given by 
\begin{equation}
U_{PMNS}=V_e^\dagger V_{\nu}
\end{equation}
where $V_e$ and $V_{\nu}$ are the diagonalization matrices for the charged leptons and neutrinos. The diagonalization of the charged lepton mass matrix $M_e$ is similar to the quark sector (Eqn.(3)). For Majorana neutrinos, the complex symmetric neutrino mass matrix is diagonalized by a unitary matrix $V_{\nu}$ (Eqn.(6)). 
Takagi factorization \cite{16} can diagonalize a general complex symmetric neutrino mass matrix without any simplifying assumption for the phases. For a complex symmetric neutrino mass matrix $M_{\nu}$ there exists a unitary matrix $V_{\nu}$ and a real non-negative diagonal matrix $M_{\nu}^{diag}$ such that 
\begin{equation}
M_\nu=V_\nu M_ \nu^{diag}V_\nu^T
\end{equation}
where the columns of $V_{\nu}$ are an orthonormal set of eigenvectors for $M_{\nu}M_{\nu}^\dagger$. The corresponding diagonal entries of $M_{\nu}^{diag}$ are non-negative square roots of the corresponding eigenvalues of $M_{\nu}M_{\nu}^\dagger$.
The information about neutrino masses and mixings can be obtained from Eqn.(33) 
\begin{equation}
\tan^2\theta_{12}=\frac{|(U_{PMNS})_{12}|^2}{|(U_{PMNS})_{11}|^2},
\end{equation}
\begin{equation}
\sin^2 2 \theta_{23} =\frac{4|(U_{PMNS})_{23}|^2|(U_{PMNS})_{33}|^2}{(1-|(U_{PMNS})_{13}|^2)^2},
\end{equation}
and
\begin{equation}
\sin\theta_{13}=|(U_{PMNS})_{13}|
\end{equation}
In order to obtain consistent values of the lepton mixing angles we have to fine tune the unknown parameters $\rho$, $\gamma$ and $\delta{'''}$. For example, for $d_e=0.93575 m_\tau$, we obtain the following mixing angles
\begin{equation}
\theta_{12}=33.45^o,
\end{equation}
\begin{equation}
\theta_{23}=47.92^o,
\end{equation}
\begin{equation}
\theta_{13}=2.035^o.
\end{equation}
The unknown parameters $\rho$, $\gamma$ and $\delta{'''}$ are fine tuned to be
\begin{eqnarray}
\rho = 1.43 \times 10^{-9}, \ \ \gamma = 2 \times 10^{16}, \  \  \delta{'''}=10.
\end{eqnarray}
We obtain a small value for $R_{\nu}=\frac{\Delta m^2_{12}}{\Delta m^2_{23}}=5.37\times10^{-4}$, which is well below the experimentally allowed range. However, it must be noted that the mixing angles and mass-squared differences have been calculated for the best fit values of the parameters $\Delta \psi$, $\Delta \phi$, $d_u$, $d_d$ given in Eqns.(27-30). However, these parameters have wide ranges and one may obtain consistent values of $\theta_{13}$ and $R_\nu$ when we consider the full ranges of these parameters. This has been illustrated explicitly in the Appendix by picking up some representative values of the parameters from within their allowed ranges to calculate the neutrino mass squared differences and the mixing angles which are in agreement with the currently available data. Our objective in using the best fit values is only meant to illustrate the methodology presented in this work. 

\section{Conclusions}
We discussed the four zero texture quark/lepton mass matrices in the context of SO(10) GUT. The integration of these two scenarios reduces the number of free parameters to one when we incorporate the known data from the quark and the charged lepton sectors. In our analysis, firstly we obtained two regions of solutions for the unknown parameter $d_e$ from the phenomenological analysis for the four zero texture lepton mass matrices.Then we used the SO(10) relations to further constrain the allowed parameter space. We successfully fitted all the unknown parameters from four zero textures and SO(10) GUT within the framework of Type-I+II seesaw mechanism. The solar/atmospheric mixing angles obtained in this analysis are in fairly good agreement with the data. Of course, one must run down the neutrino mass matrix from the GUT scale to the electroweak scale and fit with the known oscillation data. However, it is well known that the effects of RG running are negligible for mass matrices with the normal hierarchy peculiar of the texture structure considered in this work.

\textbf{\textit{\Large{Acknowledgements}}} \\
The research work of S. D. is supported by the University Grants
Commission, Government of India \textit{vide} Grant No. 34-32/2008
(SR). R. R. G. acknowledge the financial support provided by the Council for Scientific and Industrial Research (CSIR), Government of India.

\section{Appendix}
In this appendix we give the numerical values of different parameters which satisfy the experimental data using the input parameters $\Delta \psi$, $\Delta \phi$, $d_u$, $d_d$ in there allowed experimental ranges\cite{6}.
Table-I lists the numerical values of different parameters. The neutrino mixing angles and mass squared differences for these values of parameters are
\begin{equation}
\theta_{12}=32.26^o,\ \ \ \ 
\theta_{23}=45.74^o,\ \ \ \ 
\theta_{13}=12.34^o.
\end{equation}
\begin{equation}
\Delta m_{12}^{2} =7.43\times 10^{-5}eV^{2},\ \ \ \ \ 
\Delta m_{32}^{2} = 2.703\times 10^{-3}eV^{2}.
\end{equation}

\begin{table}
\begin{center}
\begin{tabular}{|c|c|}
\hline Parameter & Numerical value \\
\hline $d_u$ &0.951934 $m_t$ \\ 
\hline  $d_d$&0.932640 $m_b$ \\ 
\hline  $\Delta \phi$& $\pi/2$\\ 
\hline  $\Delta \psi$& -0.044058\\ 
\hline  $d_e$&0.890355 $m_{\tau}$ \\ 
\hline $r$ & 0.408963\\ 
\hline  $\rho$&$3\times 10^{-10}$ \\ 
\hline $\gamma$ & $2.238477 \times 10^{18}$\\ 
\hline  $\delta{'''}$& 29414\\ 
\hline 
\end{tabular}
\caption{Numerical values of parameters}
\end{center}
\end{table}


\begin{thebibliography}{99}
 
 
\bibitem{1}Paul H. Frampton, Sheldon L. Glashow and Danny
Marfatia, \textit{Phys. Lett.} \textbf{B 536}, 79 (2002), hep-ph/0201008; Zhi-zhong Xing, \textit{Phys. Lett.} \textbf{B 530},
159 (2002), hep-ph/0201151; Bipin R. Desai, D. P. Roy and Alexander R. Vaucher, \textit{Mod. Phys. Lett} \textbf{A 18}, 1355 (2003), hep-ph/0209035; S. Dev, Sanjeev Kumar, Surender Verma and Shivani Gupta,
 \textit{Nucl. Phys.} \textbf{B 784}, 103-117 (2007), hep-ph/0611313;
 S. Dev, Sanjeev Kumar, Surender Verma and Shivani Gupta,
 \textit{Phys. Rev.} \textbf{D 76}, 013002 (2007), hep-ph/0612102.
\bibitem{2} G. C. Branco, R. Gonzalez, F. R. Joaquim and T. Yanagida, \textit{Phys. Lett.} \textbf{B 562}, 265 (2003), hep-ph/0212341; Bhag C. Chauhan, Joao Pulido and Marco Picariello, \textit{Phys. Rev.} \textbf{ D 73}, 053003 (2006), hep-ph/0602084.
\bibitem{3} E. I. Lashin and N. Chamoun, \textit{Phys. Rev.} \textbf{D 78}, 073002 (2008), arXiv:0708.2423 [hep-ph];  E. I. Lashin, N. Chamoun, \textit{Phys.Rev.} \textbf{D 80}, 093004 (2009), arXiv:0909.2669 [hep-ph]; S. Dev, Surender Verma, Shivani Gupta and R. R. Gautam, \textit{Phys. Rev.} \textbf{D 81}, 053010 (2010), arXiv:1003.1006 [hep-ph]; S. Dev, Shivani Gupta and R. R. Gautam, \textit{Mod. Phys. Lett.} \textbf{A 26}, 501-514, arXiv:1011.5587 [hep-ph]; S. Dev, Shivani Gupta, Radha Raman Gautam and Lal Singh, \textit{Phys. Lett.} \textbf{B 706}, 168 (2011), arXiv:1111.1300 [hep-ph].
\bibitem{4} H. Fritzsch and Z. Z. Xing, \textit{Prog. Part. Nucl. Phys.} \textbf{45}, 1 (2000), hep-ph/9912358 and references therein.
\bibitem{5} W. Grimus, \textit{Proc. Sci.}, \textbf{HEP 186} (2005), hep-ph/0511078.
\bibitem{6} K. Matsuda and H. Nishiura, \textit{Phys. Rev.} \textbf{D 74}, 033014 (2006), hep-ph/0606142 and references therein.
\bibitem{7} Takeshi Fukuyama, Koichi Matsuda and Hiroyuki Nishiura, \textit{Int. J. Mod. Phys.} \textbf{ A 22}, 5325-5343 (2007), hep-ph/0702284; K. Matsuda, H. Nishiura and T. Fukuyama, \textit{Phys. Rev.} \textbf{D 61}, 053001 (2000), hep-ph/9906433.
\bibitem{8} S. Dev, Sanjeev Kumar, Surender Verma and Shivani Gupta, \textit{Mod. Phys. Lett.} \textbf{A 24}, 28(2009), arXiv:0810.3083 [hep-ph]
\bibitem{9}  N. Cabibbo,  \textit{Phys. Rev. Lett.} \textbf{10}, 531 (1963); M. Kobayashi and T. Maskawa, \textit{Prog. Theor. Phys.} \textbf{49}, 652 (1973).
\bibitem{10} B. Pontecorvo,  \textit{Zh. Eksp. Teor. Fiz. (JTEP)} \textbf{33}, 549 (1957); \textit{ibid.} \textbf{34}, 247 (1958);  \textit{ibid.} \textbf{53}, 1717 (1967); Z. Maki, M. Nakagawa and S. Sakata, \textit{Prog. Theor. Phys.} \textbf{28}, 870 (1962). 
\bibitem{11} C. Amsler \textit{et al}. [Particle Data Group], \textit{Phys. Lett.} \textbf{B 667}, 1 (2008).
\bibitem{12} Proceedings of NO-VE 2008, IV International workshop on "Neutrino Oscillations in Venice" (Venice, Italy, April 15-18, 2008), edited by M. Baldo Ceolin (University of Padova publication, Papergraf Editions, Padova, Italy, 2008), pages 21-28, hep-ph/0809.2936
\bibitem{13} K. Abe et al. [T2K collaboration], \textit{Phys. Rev. Lett.} \textbf{107}, 041801 (2011), arXiv:1106.2822 [hep-ex].
\bibitem{14} P. Adamson et al. [MINOS collaboration], \textit{Phys. Rev. Lett.} \textbf{107}, 181802 (2011), arXiv:1108.0015 [hep-ex].
\bibitem{15} W. Konetschny and W. Kummer, \textit{Phys. Lett.} \textbf{B 70}, 433 (1977); T. P. Cheng and L. F. Li, \textit{Phys. Rev.} \textbf{D 22}, 2860 (1980); J. Schechter and J. W. F. Valle, \textit{Phys. Rev.} \textbf{D 22}, 2227 (1980); G. Lazarides Q. Shafi and C. Wetterich, \textit{Nucl. Phys.} \textbf{B 181}, 287 (1981); R. N. Mohapatra and G. Senjanovic, \textit{Phys. Rev.} \textbf{D 23}, 165 (1981).
\bibitem{16} T. Takagi, \textit{Japan J. Math.} \textbf{1}, 83 (1923); R. A. Horn and C. R. Johnson, \textit{Matrix Analysis} (Cambridge University Press, Cambridge, England, 1990).

 \end{thebibliography}
\end{document}